\documentclass[12pt,preprint]{aastex}
\usepackage{times}
\usepackage{natbib}

\newcommand{\msun}{\mbox{${\rm M}_{\odot}$}}
\newcommand{\lsun}{\mbox{${\rm L}_{\odot}$}}

\newcommand{\simgt}{\lower.5ex\hbox{$\; \buildrel > \over \sim \;$}}
\newcommand{\simlt}{\lower.5ex\hbox{$\; \buildrel < \over \sim \;$}}
\newcommand{\BV}{Brunt-V\"ais\"al\"a\ }

\def\apj{ApJ}%
\def\apjl{ApJ}%
\def\apjs{ApJS}%
\def\apss{Ap\&SS}%
\def\aap{A\&A}%
\def\azh{AZh}%
\def\mnras{MNRAS}%
\def\solphys{Sol.~Phys.}%
\def\nat{Nature}%

\shorttitle{Short title}
\shortauthors{Montalb\'{a}n et al.}

\begin{document}

\title{Seismic diagnostics of red giants: first comparison with stellar models}


\author{J. Montalb\'an and  A. Miglio\altaffilmark{1} and A. Noels and R. Scuflaire  }
\affil{Institut d'Astrophysique et G\'eophysique de l'Universit\'e de Li\`ege, All\'ee du six Ao\^ut, 17 B-4000 Li\`ege, Belgium}

\and

\author{P. Ventura}
\affil{Osservatorio Astronomico di Roma-INAF, via Frascati 33, Monteporzio Catone, Rome, Italy}


\altaffiltext{1}{Charg\'e de Recherches of the Fonds de la Recherche Scientifique, FNRS, rue d'Egmont 5, B-1000 Bruxelles, Belgium}


\begin{abstract}
The clear detection with CoRoT and KEPLER of radial and non-radial solar-like oscillations in many red giants paves the way to seismic inferences on the structure of such stars. We present an overview of the properties of the adiabatic frequencies and frequency separations of radial and non-radial oscillation modes for an extended grid of models. We highlight how their detection allows a deeper insight into the internal structure and evolutionary state of red giants. In particular, we find that the properties of dipole modes constitute a promising seismic diagnostic tool of the evolutionary state of red-giant stars.
We compare our theoretical predictions with the  first 34 days of KEPLER data and predict the frequency diagram  expected for red giants in the COROT-exofield in the galactic center direction. 
\end{abstract}





\section{Introduction}

Red giants are cool  stars with an extended convective envelope, which can, as in main sequence solar-like stars,  stochastically excite pressure modes 
of oscillation.  Although  stochastic  oscillations were already  detected in a few red giants from ground and space observations \citep[e.g.][]{Frandsen02, Joris06, Barban07} it has been only thanks to the photometric  space mission COROT \citep{Baglin02} that an unambiguous detection of radial and non-radial modes in a large number of red-giant stars has been achieved \citep{Joris09, Hekker09, Carrier10}. That confirmation has opened the way to the seismic study of the structure and evolution of these objects that play a fundamental role in fields such as stellar age determination and chemical evolution of galaxies.
The application of \cite{KB95} theoretical scaling laws, which relate basic seismic observables (the large frequency separation -- $\Delta\nu$--  and the frequency at maximum power --$\nu_{\rm max}$) to the stellar global parameters,  allowed \cite{Mosser10}  to estimate the masses of CoRoT and KEPLER red giants (adopting a value for the effective temperature). Moreover, the combination of these scalings  with the predictions of  population synthesis models let \cite{MiglioPop09} and \cite{MiglioPop209} to  characterize the population of  CoRoT and Kepler targets. 

These results, showing that a vast amount of information can be extracted from quite   easy-to-access
seismic observables have deeply changed the perception of the predictive capabilities of asteroseismology and have strengthened its interaction with the other fields of astrophysics. However, much more information is contained in the oscillation spectra of these large number of red giants. In this Letter, we analyze  the properties of red giant adiabatic oscillation spectra and relate them with their evolutionary state.

\begin{figure*}[ht!]
\centering
\resizebox{\hsize}{!}{\includegraphics{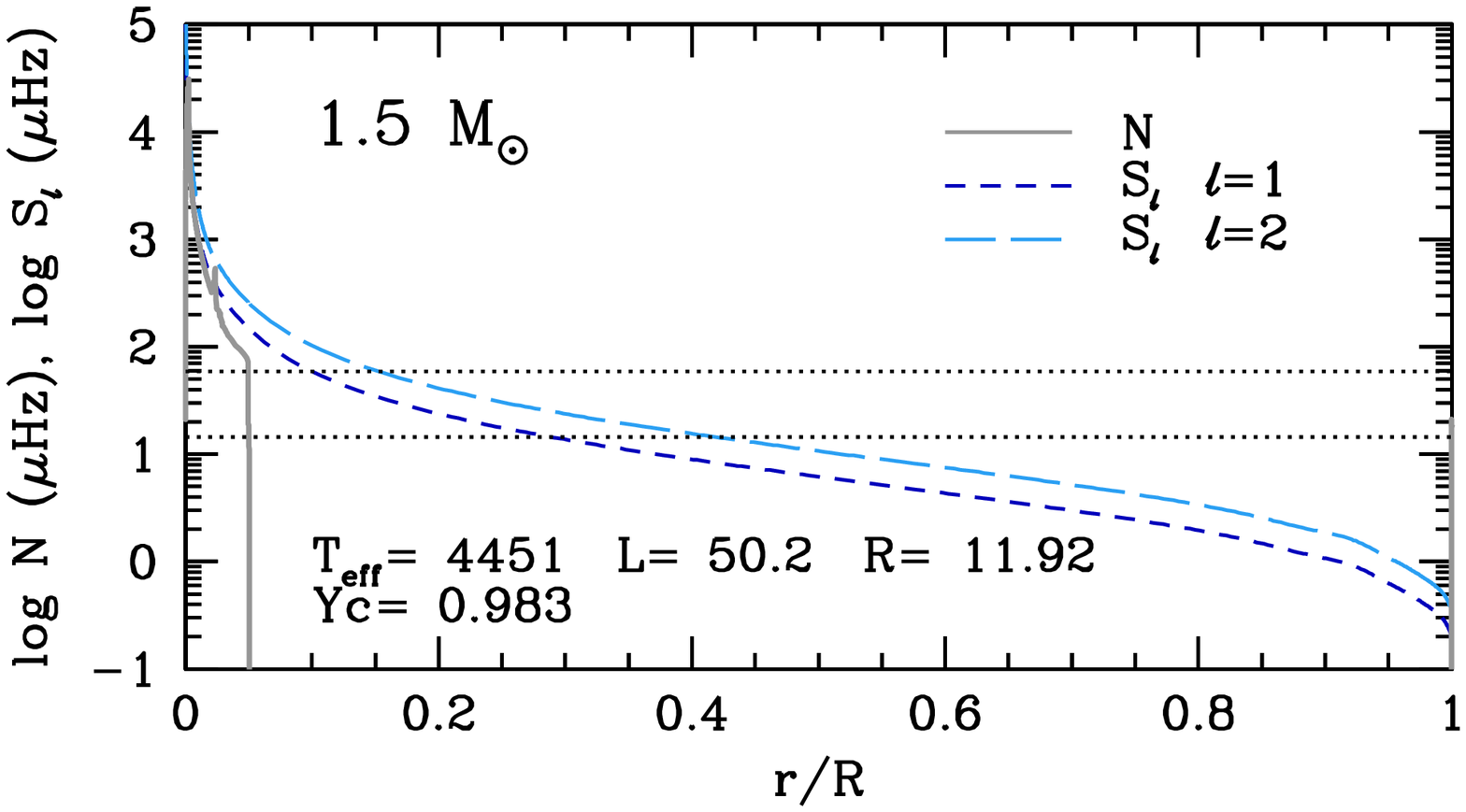}\includegraphics{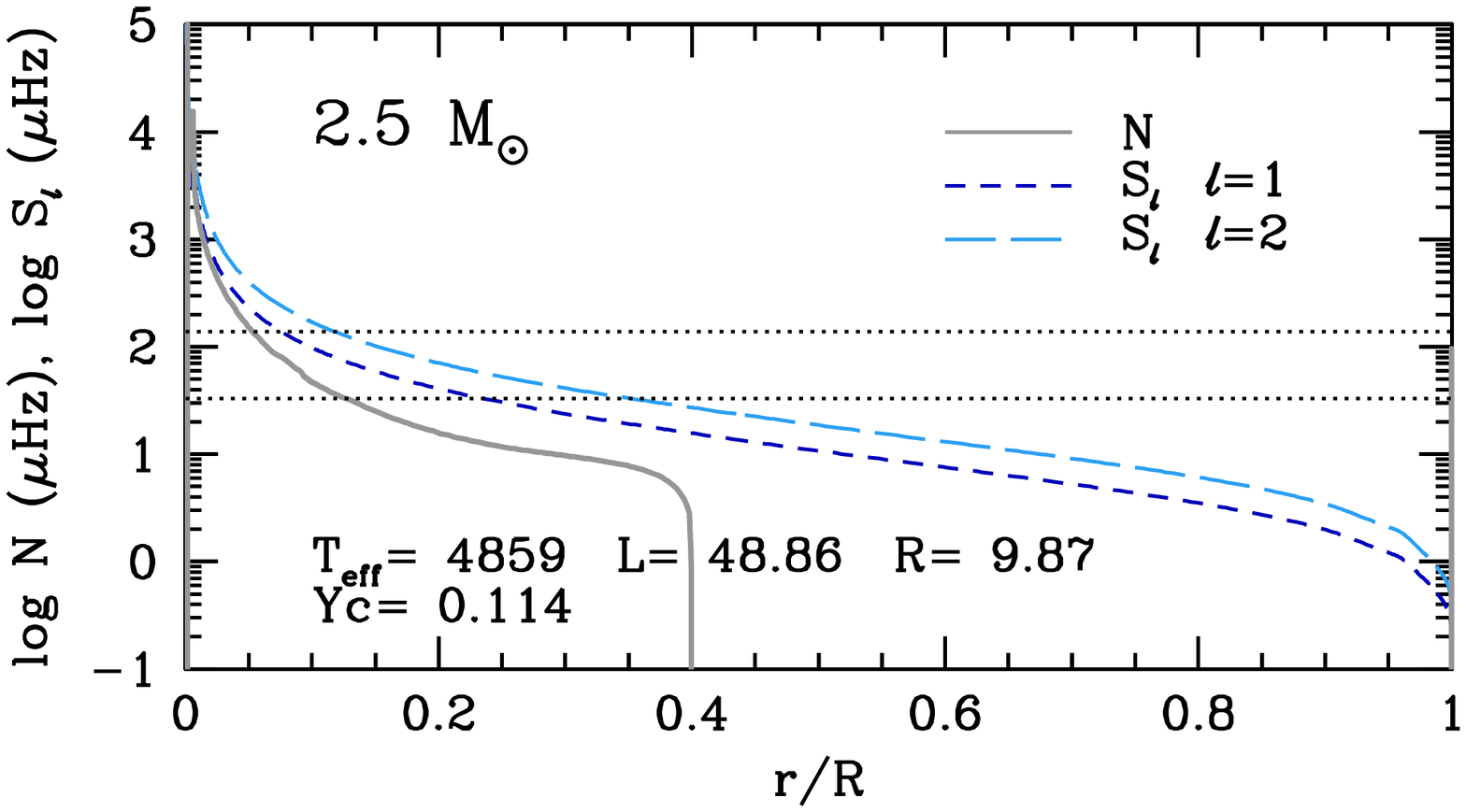}}
\resizebox{\hsize}{!}{\includegraphics{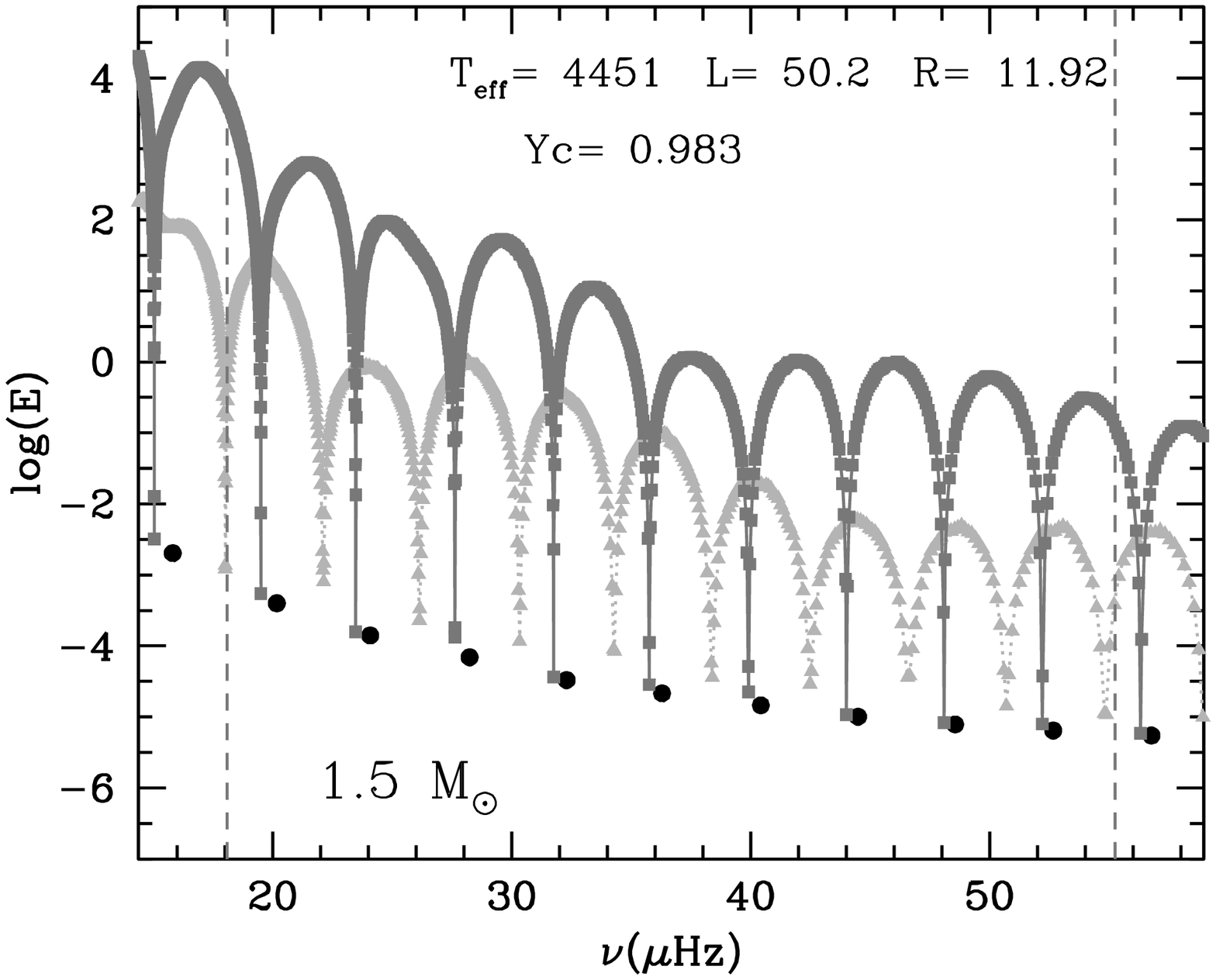}\includegraphics{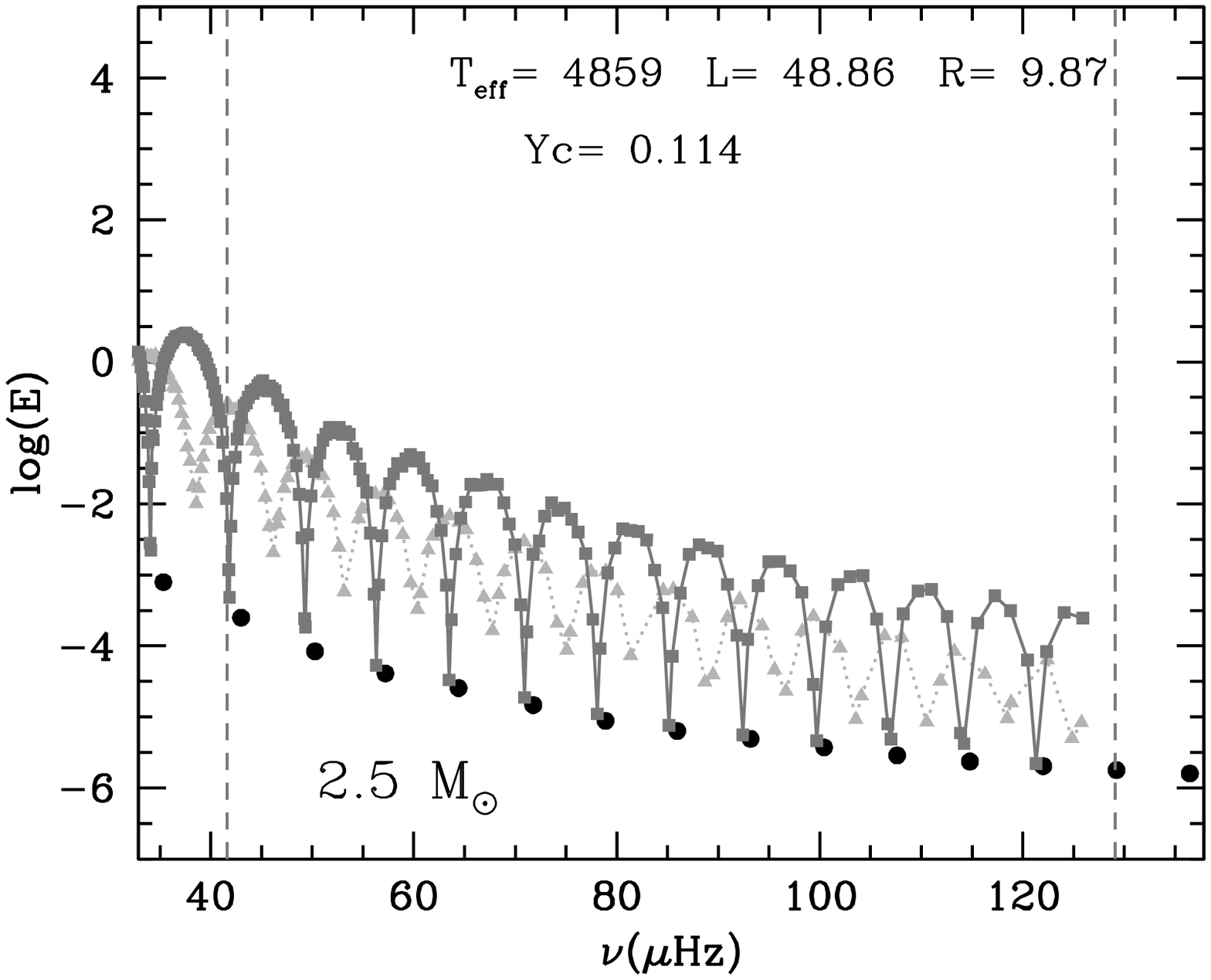}}
\vspace*{-1.75cm}
\caption{Upper panels: propagation diagrams for two models at almost the same luminosity,  1.5~\msun\  in the RGB (left) and 2.5\msun\  in central He-burning phase (right). Horizontal dotted lines limit the solar-like frequency domain for each model. Solid line is \BV\ frequency, and short- and long-dashed lines correspond to the Lamb frequency for $\ell=1$ and 2 respectively.  Lower panels: corresponding plots of  inertia as a function of frequency for $\ell$=0 (black circles), 1 (grey triangles) and 2 (dark-grey squares) modes in the solar-like oscillation domain (dashed vertical lines).}
\label{label_prop}
\end{figure*}

\section{Stellar Models}
\label{sec_models}
Stellar models were computed with the code ATON3.1 \citep[][and references therein]{Ventura08} for masses between 0.7 and 5.0 \msun\, and  different chemical compositions:  He mass fraction, $Y$=0.25 and 0.278,  and metal mass fraction, $Z$=0.006, 0.01, 0.015, 0.02 and 0.03. The energy transport in the convective regions was modeled with the classic mixing length treatment with  $\alpha_{\rm MLT}=1.6$. For a given chemical composition ($Z=0.02$, $Y=0.278$), models with  $\alpha_{\rm MLT}=1.9$ and  FST treatment of convection \citep{Canutoetal96} were also computed. The evolution of these models was followed from the pre-main sequence until the exhaustion of He in the core for models more massive than 2.3\msun, and until the helium flash for the less massive ones. The core He-burning phase for low-mass ($0.7-2.3$\msun) stars (Red clump stars) has also been followed starting from  zero age horizontal branch  models. Microscopic diffusion was not included but its effects on red giant models \citep{michaud10} are not relevant for the present study.

The evolution rate during  ascending RGB, descending  RGB and core He-burning phases is very different and strongly  depends on stellar mass. For  low-mass stars the  time spent ascending the RGB  may be comparable with that of  core He-burning of more massive stars. As a consequence, observing stars in both evolutionary stages would be equally likely. Concerning the internal structure of these  models, it is worth mentioning that  for a low-mass model (1.5~\msun, for instance) the density contrast ($\rho_c/\langle\rho\rangle$, central to mean density ratio) changes from $10^6$  at the bottom of its RGB to  $3\times 10^9$  at $\log L/L_\odot\sim 2$. Models in the core He-burning phase, on the other hand,  have a much lower value of  $\rho_c/\langle\rho\rangle$ of the order of $2\times 10^7$; moreover,  due to  the high dependence on temperature of the 3$\alpha$ nuclear reactions, they develop  a small convective core. At a given  luminosity, e.g. that of the Red Clump,  the value of $\rho_c/\langle\rho\rangle$ for a 1.5\msun\ RGB model is more than 10 times that of  a He-burning one. So different structures should imply significant effects on the oscillation properties.

\section {Adiabatic oscillation properties}

Adiabatic oscillation frequencies were computed with an Eulerian version of the code LOSC \citep{Scuflaire08} for models from the bottom of the RGB until a maximum luminosity  ($\log L/L_\odot ~ 2.2-3.2$, depending on mass) and as well during the phase  of core  He-burning. In this paper we deal with adiabatic computations and do not consider the problem of excitation and damping of solar-like oscillations in red giants \cite[][]{Dziembowski01, HoudekGough02, Dupret09}. We use the scaling laws \citep[][]{Brownetal91, KB95} to derive  the frequency domain in which solar like oscillations are expected, and the value of the mode inertia as an estimate of the expected mode amplitude \citep[][]{JCD04}.  We search oscillation modes with angular degree $\ell=0$, 1, 2 and 3\footnote{$\ell=3$ only for $Z=0.02$, $Y=0.278$, $\alpha_{\rm MLT}=1.9$}  in the domain of frequencies defined by  an interval around $\nu_{\rm max}$ \citep[Eq. (10) of][]{KB95}. The width of the solar-like frequency domain is taken to be 20\% larger than the  difference between the acoustic cutoff frequency in the stellar atmosphere  and $\nu_{\rm max}$. 

\begin{figure}[ht!]
\centering
\resizebox{\hsize}{!}{\includegraphics{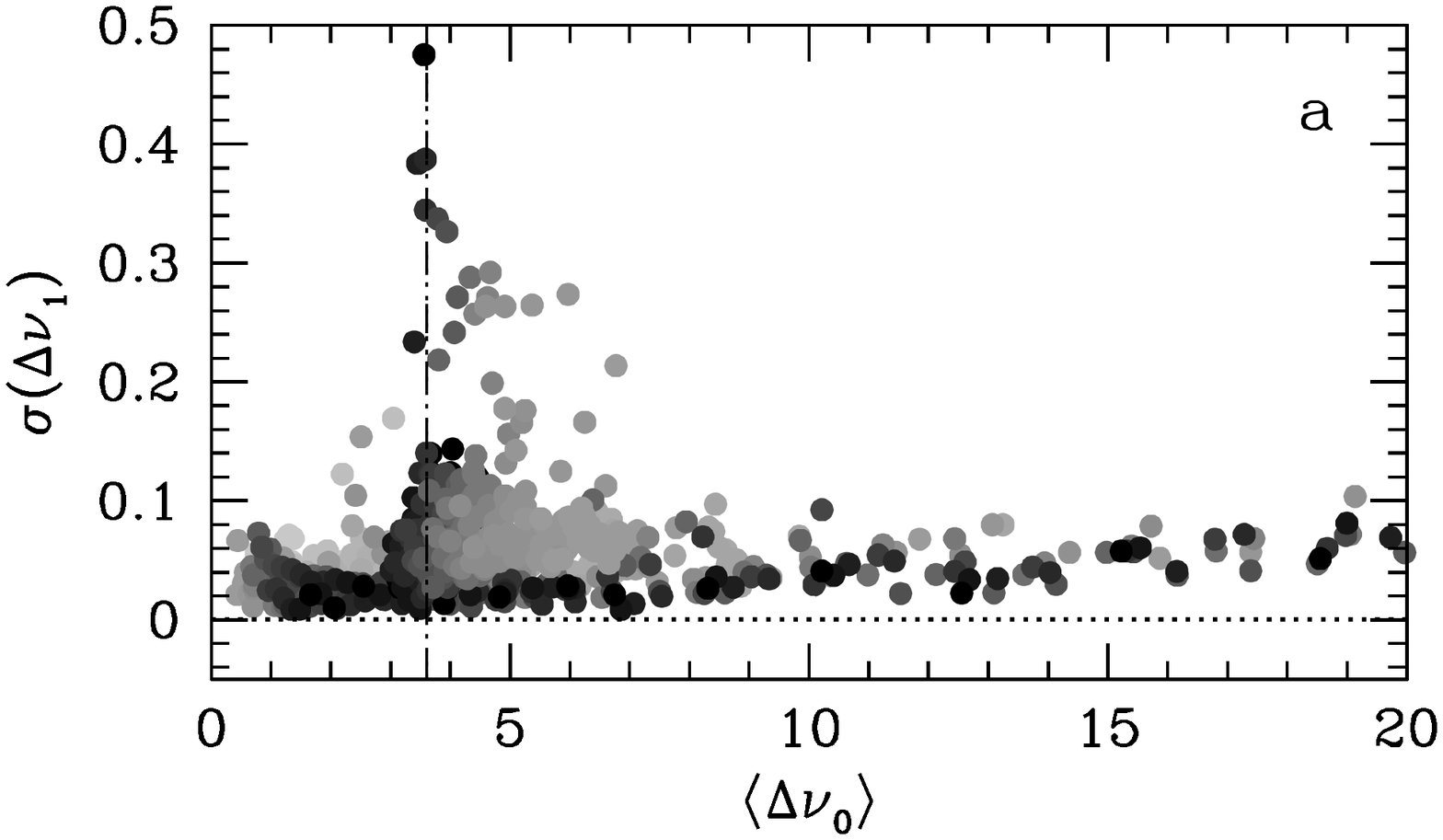}}
\resizebox{\hsize}{!}{\includegraphics{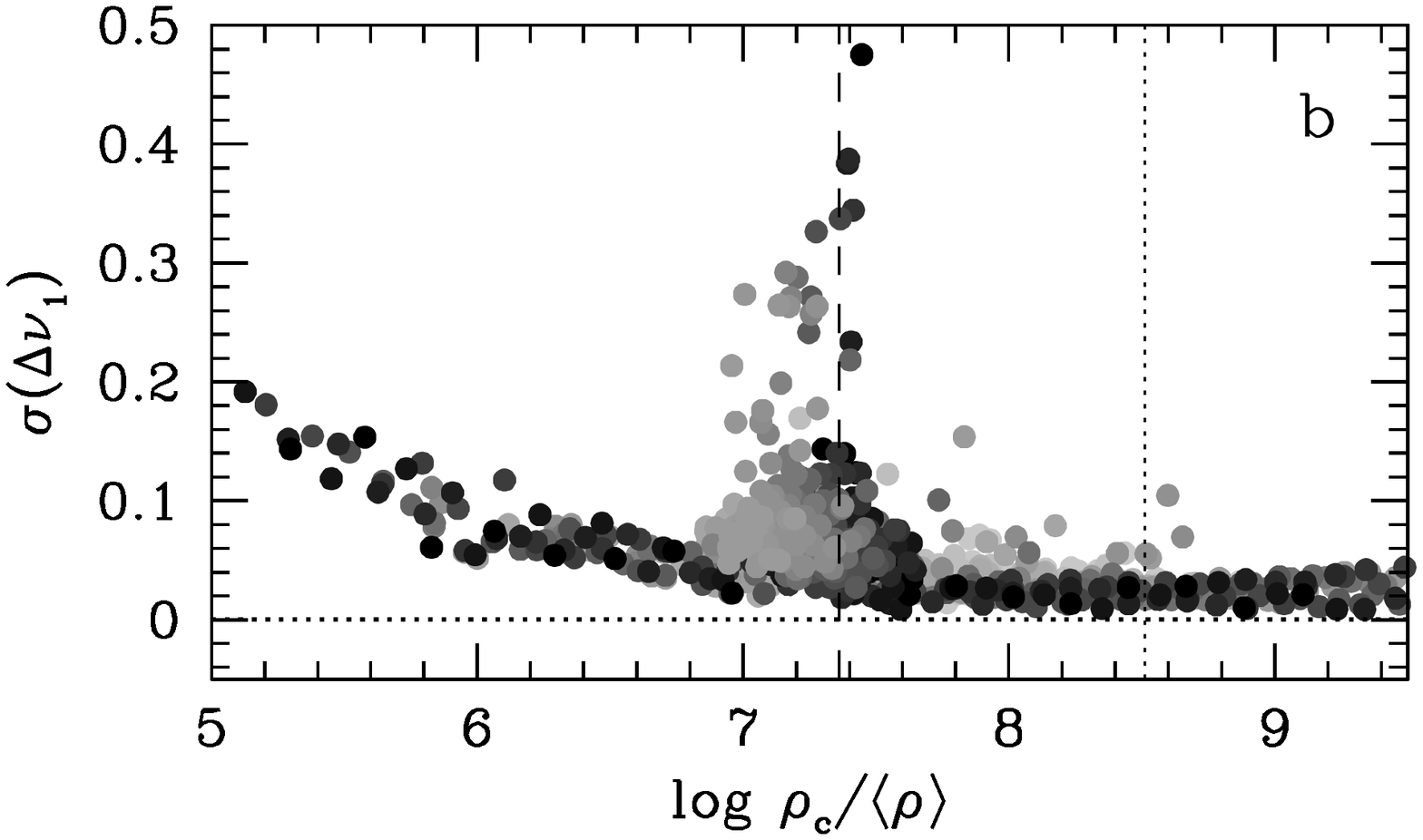}}
\caption{Mean squared deviation with respect to the main value of $\Delta\nu_{\ell=1}$, for RGB and core He-B models between 0.7\msun\ (darkest symbols) and 5.0\msun\ (lightest symbols), as a function of mean large separation for radial modes (a), and as a function of the density contrast (b). Vertical dotted and dashed lines correspond to 1\msun\ models at $\log L/L_{\odot} \sim 1.68$ in the RGB and core He-B phases respectively.}
\label{label_sigma}
\end{figure}

\begin{figure}[ht!]
\centering
\resizebox{\hsize}{!}{\includegraphics{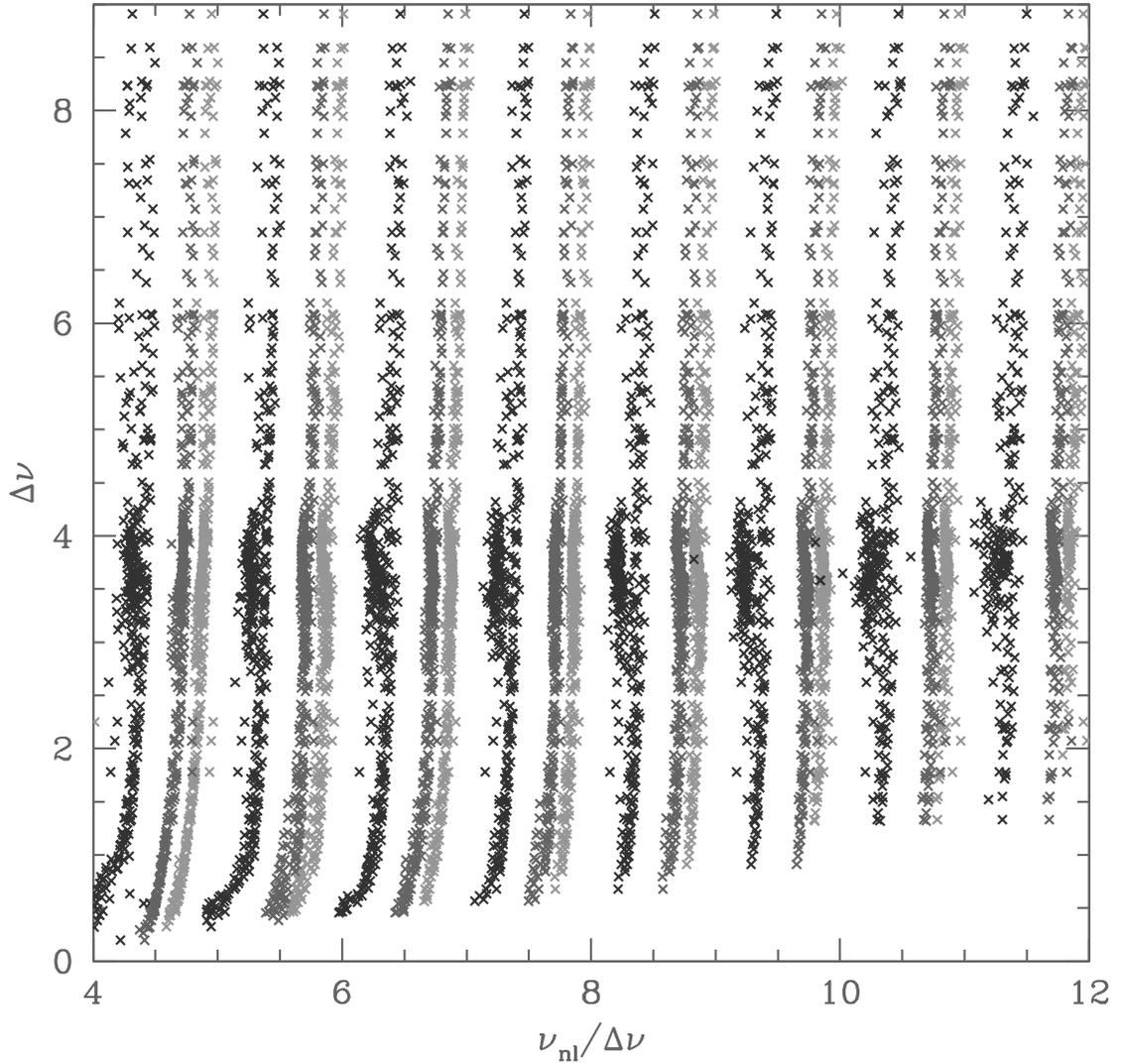}}
\caption{ Theoretical oscillation pattern expected for $\ell=0$ (light grey), 1 (black) and 2 (dark-grey) modes from the red giant population in the CoRoT-exofield in the galactic center direction. In ordinate the large frequency separation for radial modes, and in abscissa the frequency of ($n,\ell$) modes are normalized by the large separation.}
\label{label_mosser}
\end{figure} 

\begin{figure*}[ht!]
\centering
\resizebox{\hsize}{!}{\includegraphics{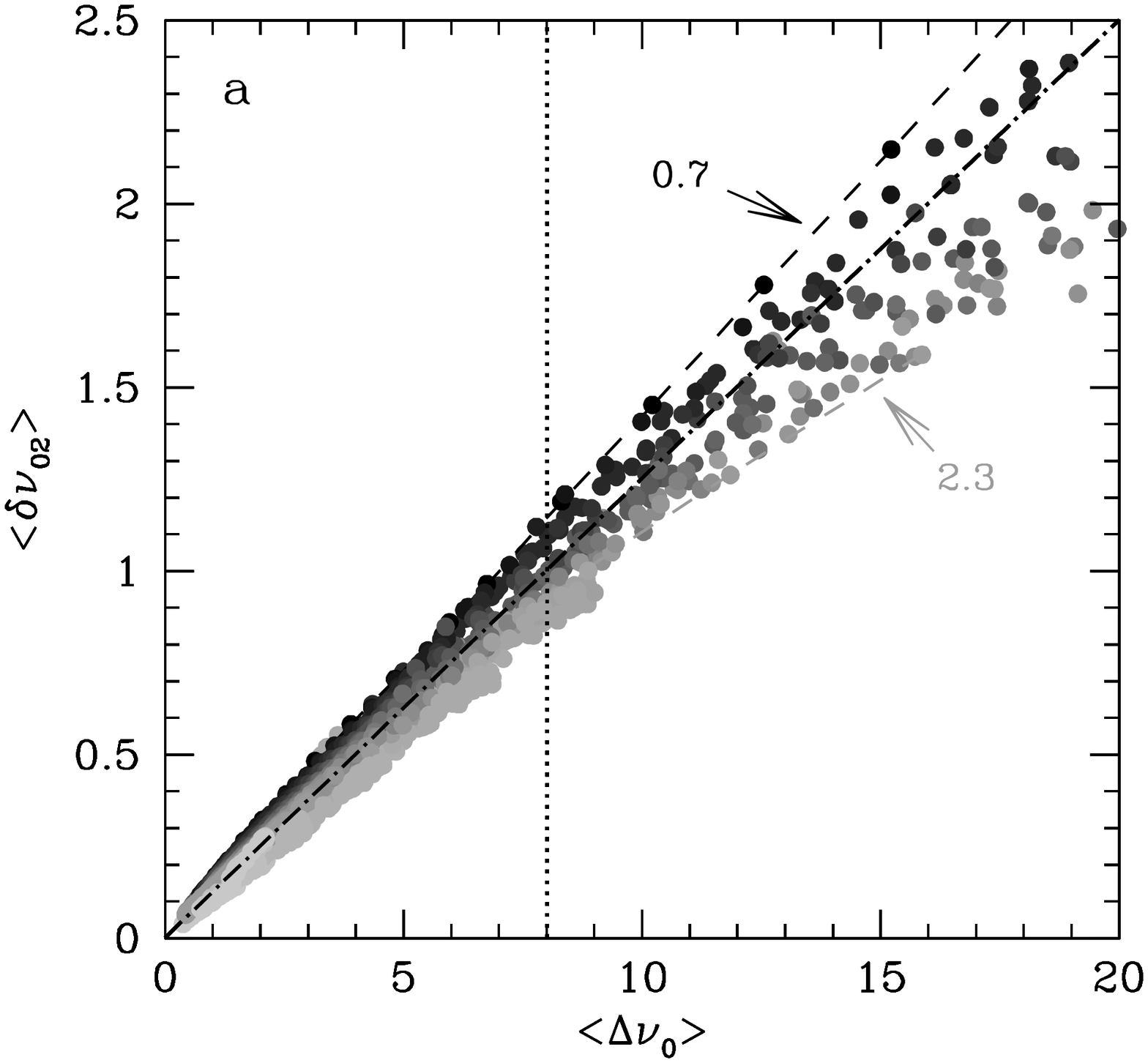}\includegraphics{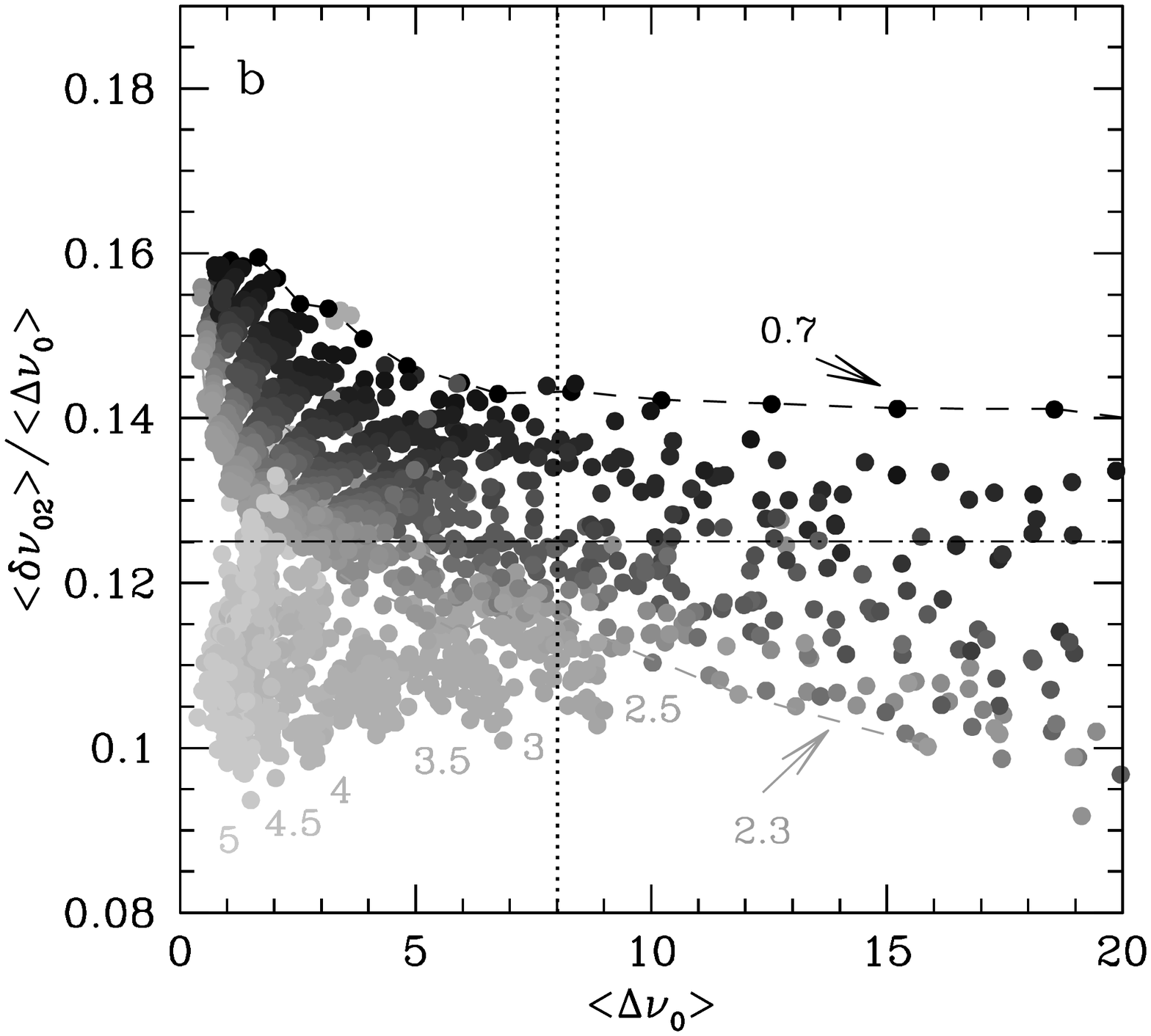}\includegraphics{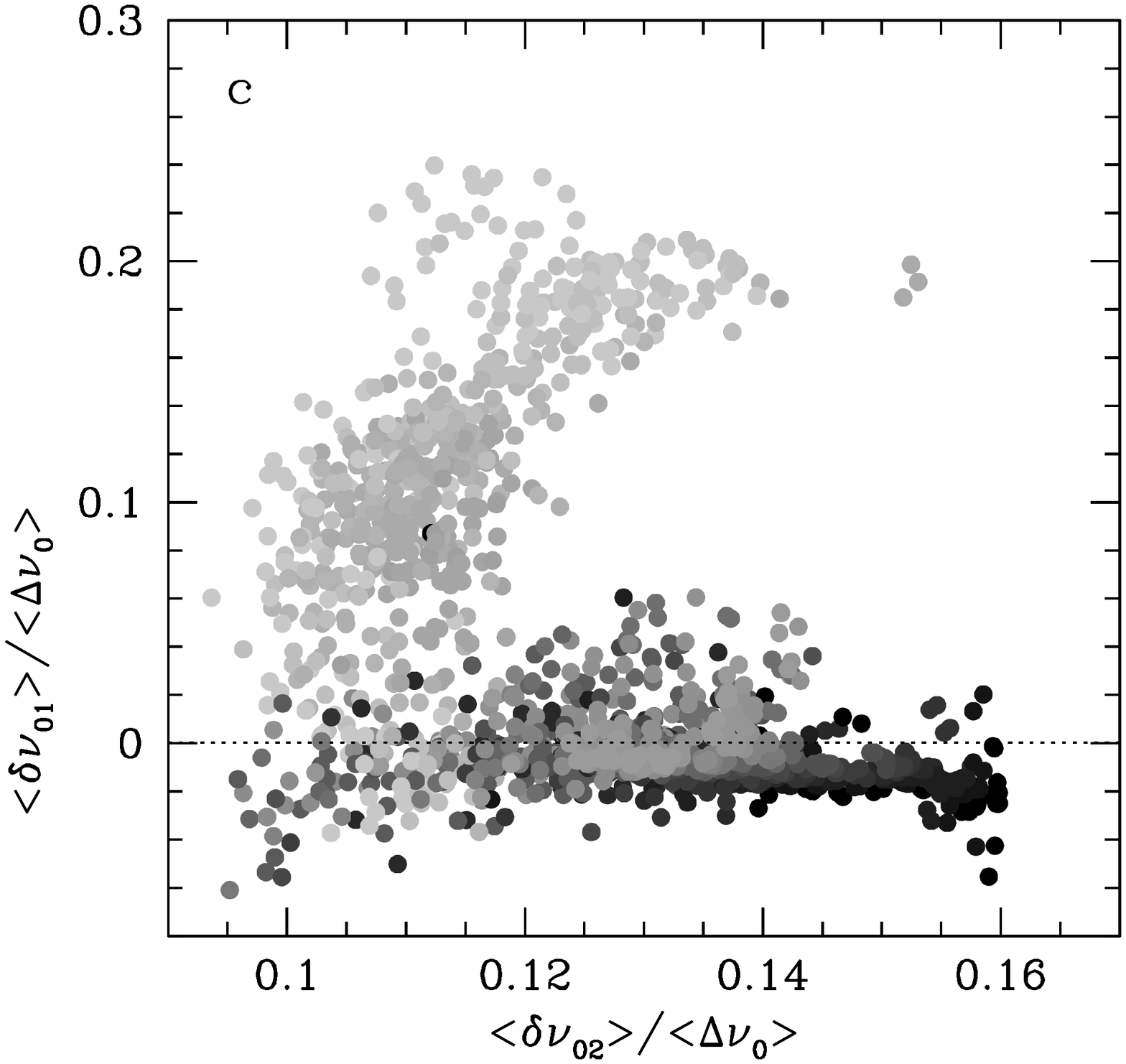}}
\resizebox{\hsize}{!}{\includegraphics{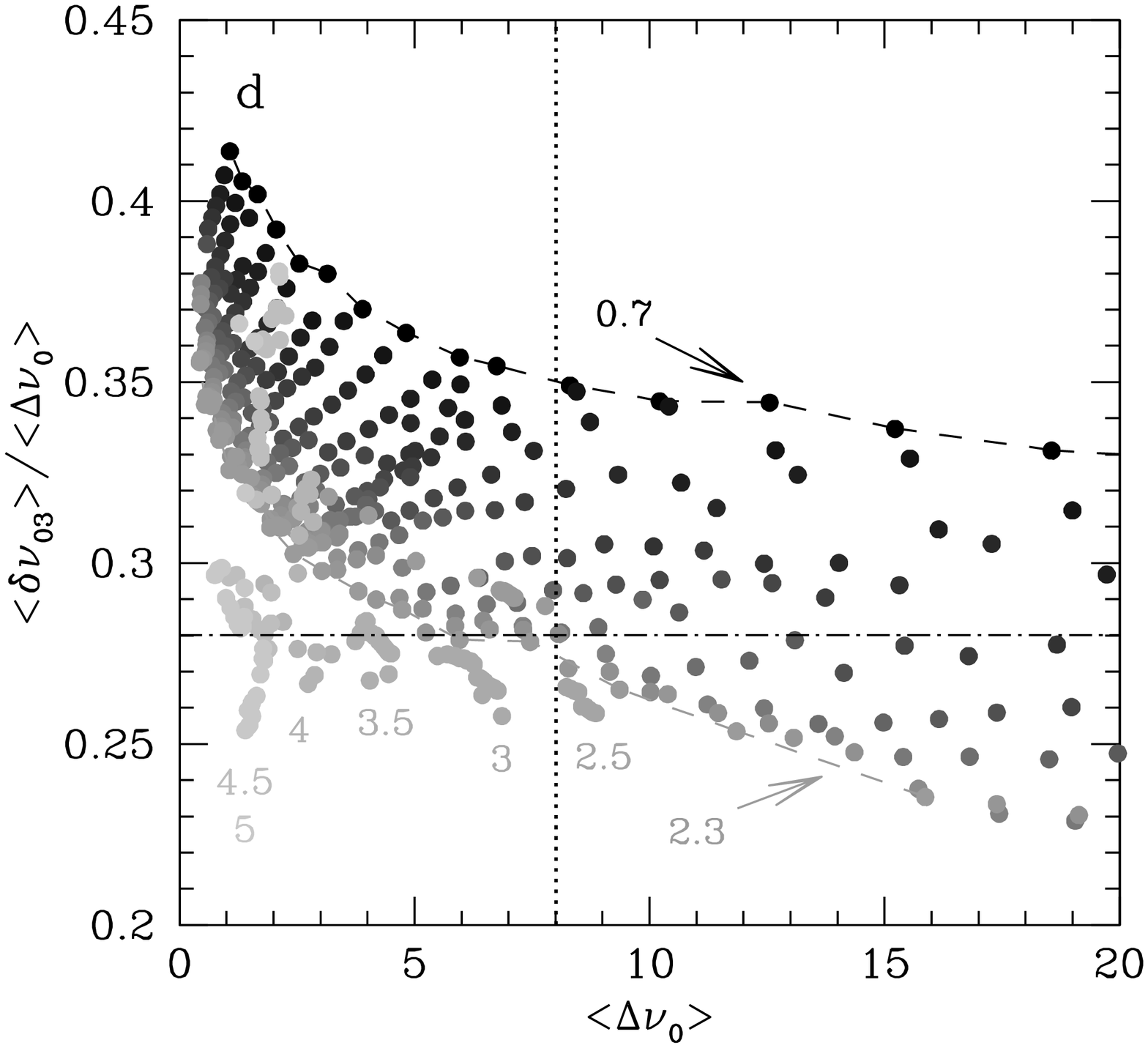}\includegraphics{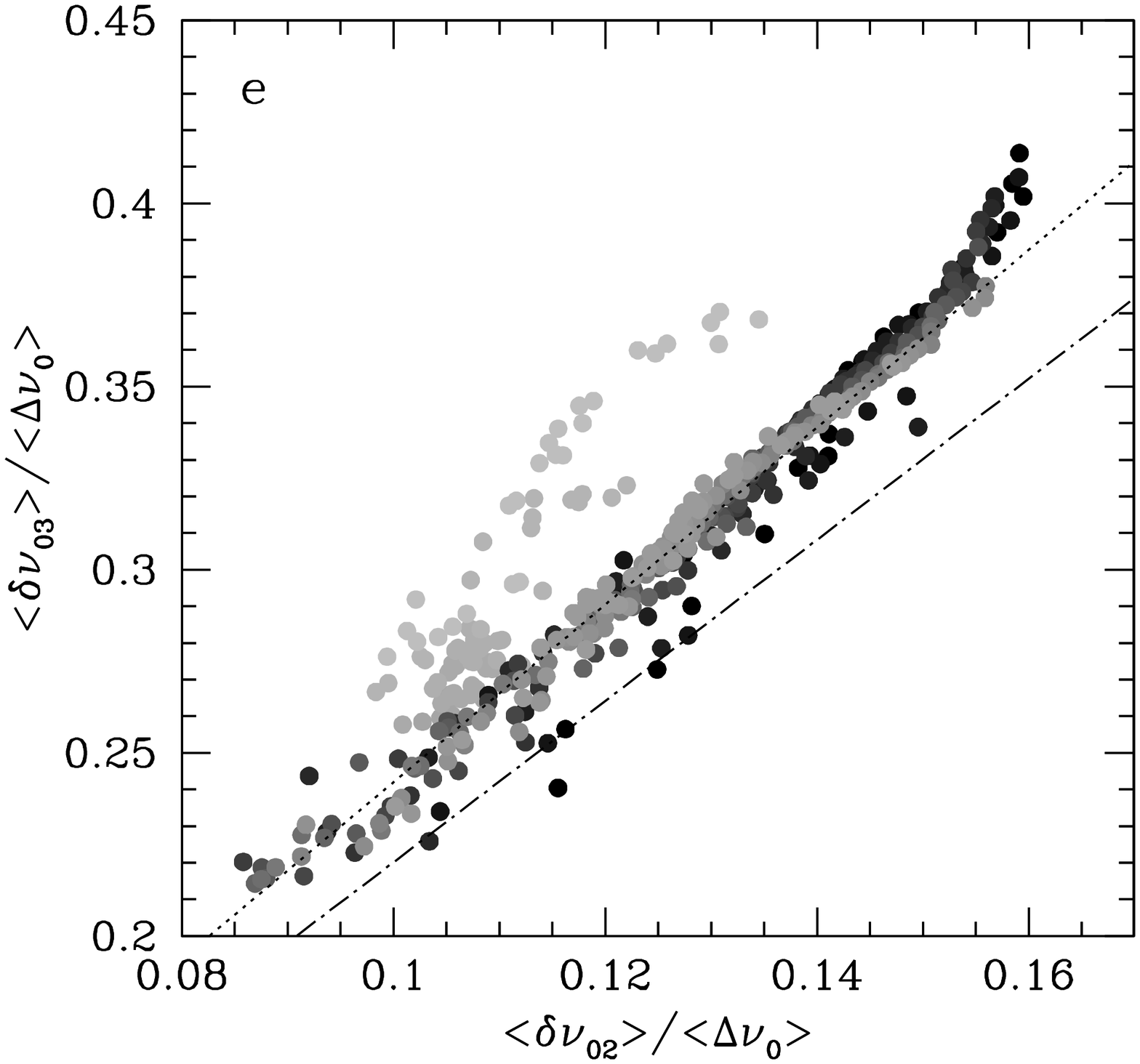}\includegraphics{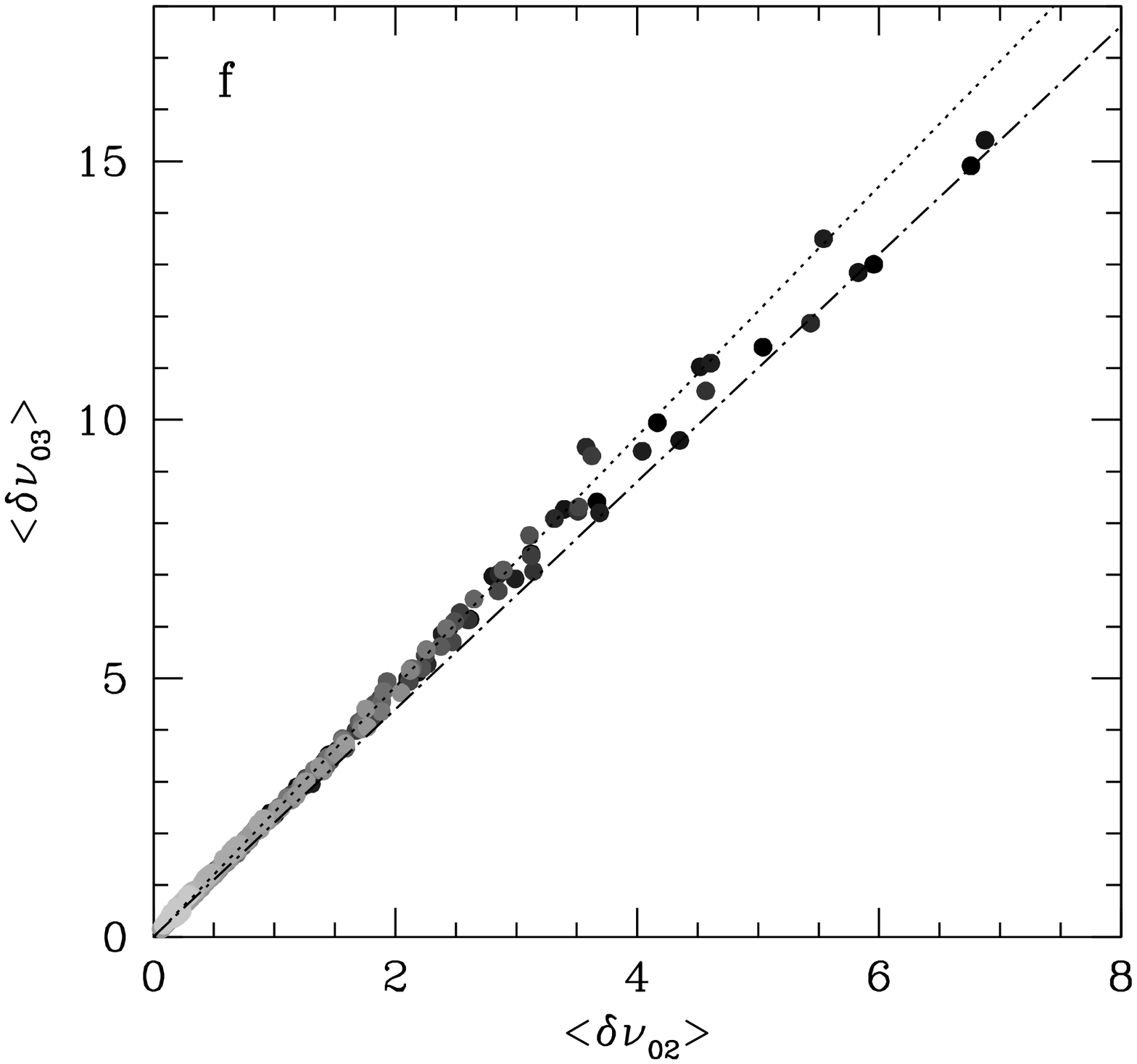}}
\caption{Small frequency separations for models with masses between 0.7 and 5~\msun\ (each grey level corresponds to a different mass, from the darkest for 0.7\msun, to the lightest for 5\msun), $Z$=0.006, 0.015, 0.02 and 0.03, $Y$=0.025 and 0.278, $\alpha_{\rm MLT}=1.6$ and 1.9. Dotted vertical line indicates the minimum $\langle\Delta\nu\rangle$ from the first 34-d of KEPLER data, and the dot-dashed lines correspond to the linear fits for these data from  \cite{Beddingetal10}. a)  $\langle\delta\nu_{02}\rangle$ as a function of $\langle\Delta\nu_0\rangle$, for  low-mass models (0.7-2.3,\msun) from the bottom of the RGB up to  $\log L/L_\odot ~ 2.5-3$, and 2.5-5\msun\ models in the He-burning phase ($Y_c$ between 0.9 and 0.1).  b) Same as a) but for the $\langle\delta\nu_{02}\rangle/\langle\Delta\nu_0\rangle$. c) $\langle\delta\nu_{01}\rangle/\langle\Delta\nu_0\rangle$ as a function of  $\langle\delta\nu_{02}\rangle/\langle\Delta\nu_0\rangle$ for the same models as in a) and b). Note the high concentration of RGB models with small and negative value of $\langle\delta\nu_{01}\rangle/\langle\Delta\nu_0\rangle$. d) $\langle\delta\nu_{03}\rangle/\langle\Delta\nu_0\rangle$ {\it vs.} $\langle\Delta\nu_0\rangle$  for models with $Y=0.278$, $Z=0.02$, and $\alpha_{\rm MLT}=1.9$. e) $\langle\delta\nu_{03}\rangle/\langle\Delta\nu_0\rangle$ {\it vs.} $\langle\delta\nu_{02}\rangle/\langle\Delta\nu_0\rangle$ and models used in d). f) $\langle\delta\nu_{03}\rangle$ as a function of $\langle\delta\nu_{02}\rangle$ and models in d).}
\label{label_d02}
\end{figure*}

The  properties of oscillation modes depend on the behaviour of the \BV\ ($N$) and Lamb ($S_{\ell}$) frequencies. In red-giant models, $N$ reaches huge values in the central regions and therefore the frequency of gravity modes (g-modes)  and their number by frequency interval ($n_g$) increase with respect to main sequence models. On the other hand, the low mean density makes  the frequency of pressure modes (p-modes) to decrease. All that leads to an oscillation spectrum for red-giants where, in addition to   radial modes, one finds a large number of non-radial modes with mixed g-p properties. The dominant character of these non-radial modes depends on the separation between gravity and acoustic cavities, and may be estimated from the value of the normalized  mode inertia ($E$) \citep[see e.g.][ and references therein]{JCD04}. Therefore, some non-radial modes may be well trapped in the acoustic cavity and behave as p-modes presenting a  mode inertia close to that of radial modes, while modes with strong mixed g-p character have larger $E$ value.  Hereafter, we will use the term p-modes in quotation marks to refer to mixed modes with a dominant p-character.

 In Fig.~\ref{label_prop} we present, in top panels, the $\ell=1$, 2  propagation diagrams for a RGB 1.5~\msun\ model (left) and for a core He-burning (He-B) model of 2.5\msun\ (right).  In the bottom panels we plot the variation of the mode inertia with frequency for radial and non-radial  modes ($\ell=1,\,2$). As mentioned above, the RGB model is ten times more centrally condensed than the He-B one. The huge difference in  density between the central region and the convective envelope entails a high potential barrier between the acoustic and the gravity cavities reducing the interaction between p and g modes. As a consequence,  we find  for RGB models that $\ell=1$ modes  with $E_{\ell=1}\approx E_{\ell=0}$  are quite regularly spaced in frequency. For He-B ones,  the coupling between these cavities is more important and $\ell=1$ modes are mixed modes with $E_{\ell=1}> E_{\ell=0}$. Nevertheless, $E_{\ell=1}$ presents  still a minimum value for modes between two consecutive radial ones showing a  somewhat regular pattern. Even if the $E$ value is larger than that corresponding to radial modes  we can still consider those modes, based on the value of $E$, as observable ``p-modes''. For $\ell=2$ modes, the coupling  between the g- and p-cavities is smaller  than for $\ell=1$ and hence the trapping more efficient.  Therefore, independently of the central  condensation of the model,  a spectrum of regularly spaced $\ell=2$ ``p-modes''  with $E_{\ell=2}\approx E_{\ell=0}$ is expected. Finally, note that  the turning points for acoustic modes ($tp_\ell$ defined as the point where $\nu_{\rm max}=S_{\ell}$) are inside the convective envelope for the RGB model and in the radiative region for the He-B one.

In the asymptotic theory for p-modes \citep{vandakurov67, tassoul80, gough86} the frequencies of two modes of same degree and consecutive order are separated by a constant value $\langle\Delta \nu\rangle$  which is  approximately  independent of $\ell$ for low-degree modes, and related to the mean density of the star. Of course, the asymptotic theory is no longer valid for mixed modes and in regions with rapidly varying  physical quantities, nevertheless  the modes partially or well trapped in  the  acoustic cavity  with a dominant p-character (``p-modes'') show such regular pattern. We select then,  as ``p-modes'' of degree $\ell$, the modes with  the  minimum inertia  between  two consecutive radial modes and use them to compute the large and small frequency separations and to analyze their dependence on the stellar parameters and evolutionary state.

\subsection{Large Separation: $\Delta\nu$}

The mean value of the large frequency separation ($\langle\Delta\nu_{\ell}\rangle=\langle\nu_{n,\ell}-\nu_{n-1,\ell}\rangle$ averaged over the radial order $n$) decreases as the star ascends the RGB with a denser and denser core and a more and more diffuse envelope, but it is not possible  to distinguish, on the basis of the $\langle\Delta\nu\rangle$  value alone, among different evolutionary states, ascending RGB, descending RGB, or core He-burning.   Nevertheless, an indirect information about the evolutionary state is provided not by the average value $\langle\Delta\nu\rangle$, but by the deviation of $\Delta\nu$ as a function of frequency with respect to its mean value ($\sigma(\Delta\nu_{\ell})$). Radial and  $\ell=2$ ``p-modes'', as  mentioned above, show a very regular pattern, and  the mean quadratic deviation of $\Delta\nu(\nu)$ with respect to its mean value over the solar-like frequency domain ($\sigma(\Delta\nu_{\ell})$) is always smaller than 5\% for all the evolutionary states and masses considered.  On the contrary, $\sigma(\Delta\nu_1)$ strongly depends on the evolutionary state and  while its value remains small for more concentrated models, it may get values as large as 50\% for core He-burning ones.

Fig.~\ref{label_sigma} shows  how the scatter of $\ell=1$ modes depends on $ \Delta\nu$ (that is on $\langle \rho \rangle$) and on the density contrast. The scatter of $\ell=1$ frequencies decreases as the luminosity increases from the bottom of the RGB (low $\rho_c/\langle\rho\rangle$, or highest $\langle\Delta\nu\rangle$).  As the star goes up the RGB,  the $\ell=1$ modes are better trapped in the acoustic cavity and the spectra of dipole modes are more regular. These results are consistent with the  observational results obtained by \cite{Beddingetal10} for the first 34 days of KEPLER observations:  radial and quadruple modes for a sample of 50 low luminosity red giants ( $< 30$\lsun\ from scaling laws) show a low scatter, while dipole modes present a  significantly larger one.
 The large dispersion shown in Fig.~\ref{label_sigma}  around $\langle \Delta\nu_0\rangle \sim 4\,\mu$Hz and $\rho_c/\langle\rho\rangle \sim 2 \times 10^7 $ corresponds to models  burning He in the core. Following the core expansion, the external convective zone recedes, increasing the coupling  between gravity and acoustic cavities (see. Fig.~\ref{label_prop}) and the mixed character of oscillation modes.  The spectrum of ``trapped'' $\ell=1$ modes is then much less regular. It is worth pointing out  that, for a given mass (1~\msun, for instance) at Red-Clump luminosity, while RGB and core He-burning models have the same value of $\langle \Delta\nu_0\rangle$,  $\rho_c/\langle\rho\rangle$ may differ by two orders of magnitude.
 
The distributions of $ \Delta\nu$ and $\nu_{\rm max}$ for  the CoRoT-exofield red giant sample show  a single dominant peak located at  $\sim 4\mu$Hz and $\sim 30\mu$Hz respectively \citep[][]{Hekker09} which was interpreted by \cite{MiglioPop09} as being consistent with a population of red giants dominated by Red-Clump stars.  According to population synthesis simulations done with  TRILEGAL \citep[][]{girardietal05}  for  the CoRoT field in the galactic center direction, 70\% of the stars at the Red-clump luminosity are low-mass stars in the core He-burning phase, and therefore, with oscillation spectra significantly different from those of the other 30\% of stars that, at the same luminosity, are in the RGB phase.
In Fig.~\ref{label_mosser} we plot the ``ensemble oscillation pattern''  that we expect for the CoRoT red giant population in the galactic center direction. This diagram was obtained by taking the theoretical adiabatic spectra corresponding to the stellar models whose stellar parameters are closest to those resulting from population synthesis computations. Note the large scatter of $\ell=1$ modes in the region around $\langle \Delta\nu \rangle \sim 4\,\mu$Hz, and the regular pattern of $\ell=0$ and 2 modes.  Both characteristics are in good agreement with the observational results found by \cite{Mosser10b}.

\begin{figure}[ht!]
\centering
\resizebox{\hsize}{!}{\includegraphics{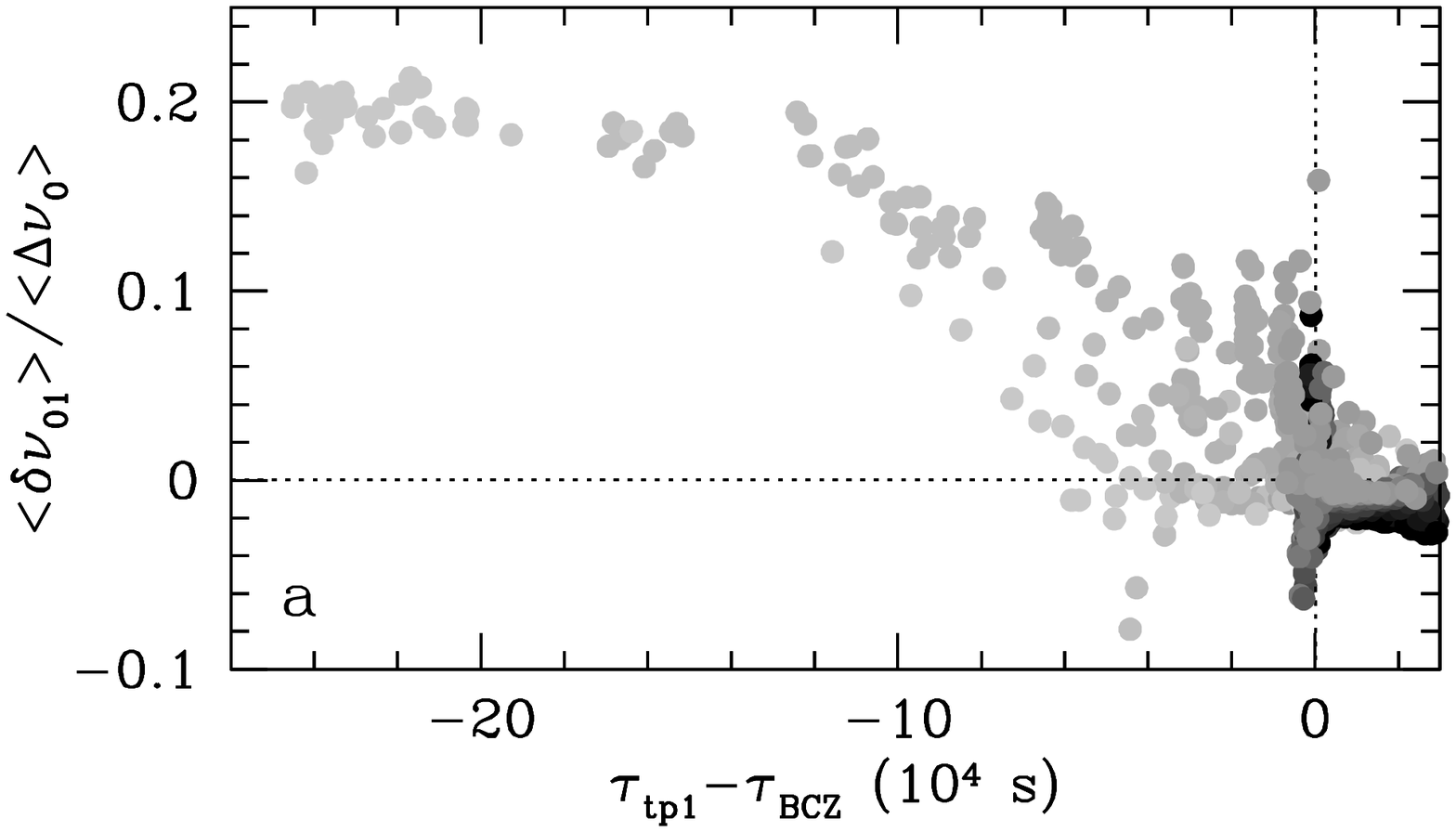}}
\resizebox{\hsize}{!}{\includegraphics{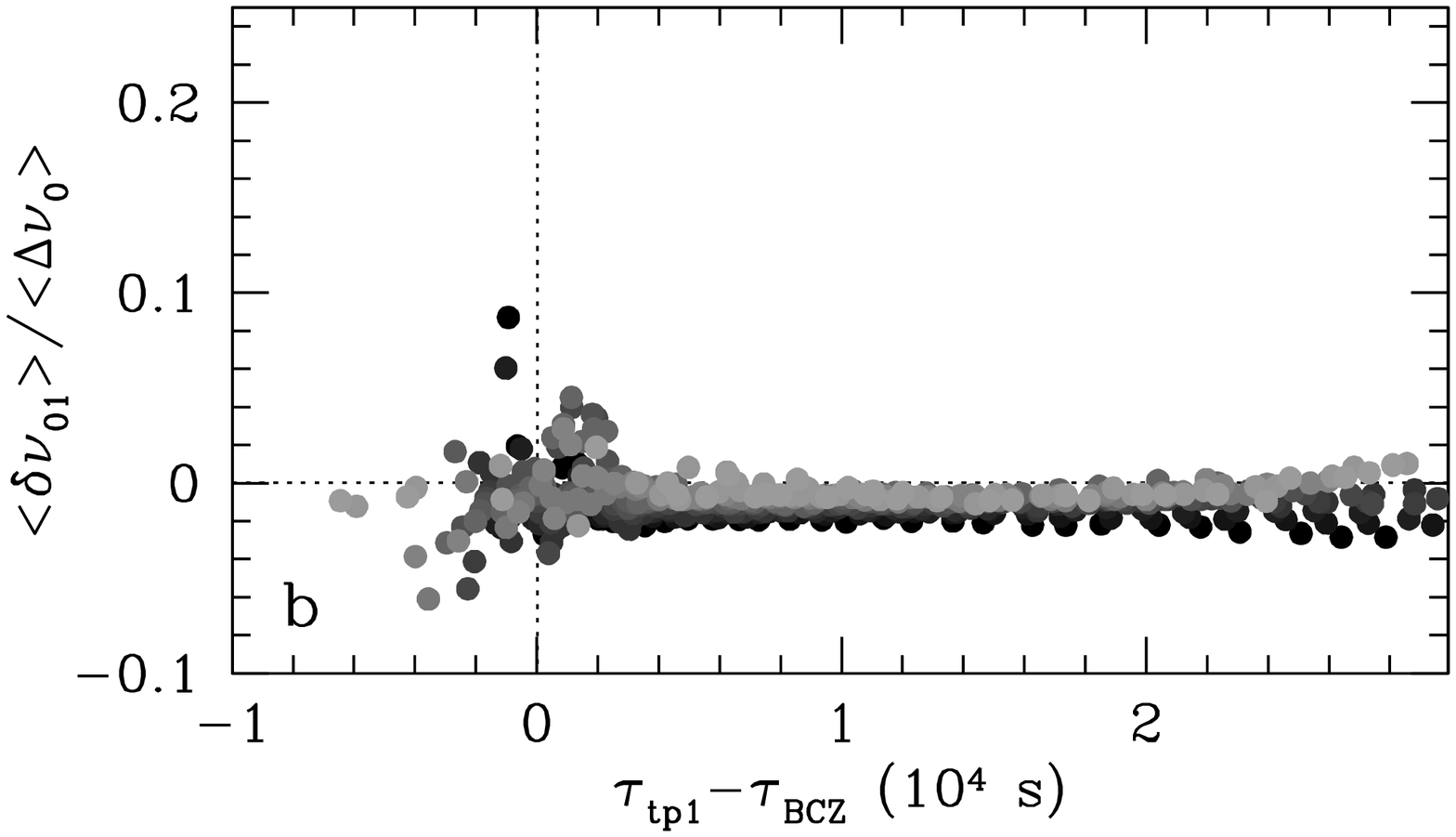}}
\caption{a) $\langle\delta\nu_{01}\rangle/\langle\Delta\nu_0\rangle$ as a function of the distance in  acoustic radius between the bottom of the convective envelope $\tau_{\rm BCZ}$  and the $\ell=1$ turning point $\tau_{\rm tp1}$  for models with i $Z=0.02$, $Y=0.278$, and $\alpha_{\rm MLT}=1.9$ and masses between 0.7 and 5~\msun.  b) zoom for low mass models (0.7--2.3\msun) in the RGB. Same grey level criteria as in Fig.4.}
\label{label_d01L}
\end{figure}

\subsection{Small separations: $\delta\nu_{02}$, $\delta\nu_{03}$ and $\delta\nu_{01}$}

According to the asymptotic theory, the  mean small frequency separation ($\langle\delta\nu_{02}\rangle=\langle\nu_{n0}-\nu_{n-1\,2}\rangle$) is  related to the behaviour of the sound speed ($c$) mostly in the central regions, and hence to the stellar evolutionary state. The representation of $\langle\delta\nu_{02}\rangle$ vs. $\langle\Delta\nu\rangle$ is in fact considered as a seismic diagnostic diagram allowing to derive stellar mass and age for main-sequence solar-like pulsators \citep{jcd88}. The corresponding seismic diagram for red-giant models is drawn in Fig.~\ref{label_d02}.a  which shows a linear  dependence of $\langle\delta\nu_{02}\rangle$ on $\langle\Delta\nu_0\rangle$, with a slope that increases as the mass decreases. In Fig.~\ref{label_d02}.b we plot the normalized quantity  $\langle\delta\nu_{02}\rangle/\langle\Delta\nu_0\rangle$ that in main-sequence stars is known to depend mostly on  central physical conditions \citep{RoxburghVorontsov03}.  It is worth noticing that we plotted  $\langle\delta\nu_{02}\rangle$ for models with different chemical compositions and convection treatment, nevertheless, a predominant  dependence on mass and radius appears. For a given mass  $\langle\delta\nu_{02}\rangle/\langle\Delta\nu_0\rangle$ increases with density contrast i.e. as the star expands with smaller values of $\langle\nu_0\rangle$, decreases as the mass increases, and  does not change significantly during the core He-burning phase. 

In Fig.~\ref{label_d02}.a and b  a vertical dotted line indicates the lower limit of $\langle\Delta\nu\rangle$ measured from the first 34d of KEPLER mission,  and  the dot-dashed line corresponds to the fit  $\langle\delta\nu_{02}\rangle=0.122\langle\Delta\nu_0\rangle$ proposed by \cite{Beddingetal10} for those observations.  The comparison between our figures and  Fig.4 in \cite{Beddingetal10} indicates that the predictions from  theoretical models of low-mass stars (1-1.5\msun) in the low luminosity part of
the ascending RGB are consistent with  observational data, as the scaling based on $\nu_{\rm max}$ and $\Delta\nu$ also suggests. In that paper, the authors also report  the detection of $\ell=3$ modes, and because of  the scatter of $\ell=1$ mode  frequencies, they suggest to use a new small separation $\delta\nu_{03}(n)=0.5\,(\nu_{0\,n-1}-2\,\nu_{3\,n-2}+\nu_{0\,n})$  instead of the classic $\delta\nu_{13}$.  Note that the observational data provide a ratio $\delta\nu_{03}/\delta\nu_{02}\,\sim\,2.2$ \citep{Beddingetal10} instead of 2 predicted by the asymptotic theory. In the bottom panels of Fig.~\ref{label_d02} we plot the theoretical results for $\delta\nu_{03}$ and its relation with $\Delta\nu_0$ and $\delta\nu_{02}$.  \cite{Beddingetal10}  results are represented by a dash-dotted line while the dotted line corresponds to our fit $\delta\nu_{03}\sim 2.42\, \delta\nu_{02}$. The dependence of $\delta\nu_{03}$ on stellar mass and radius is similar to that of  $\delta\nu_{02}$, therefore no additional information should be expected from    $\delta\nu_{03}$.

The small frequency separation $\delta\nu_{01}(n)=0.5\,(\nu_{0n}-2\,\nu_{1,n}+\nu_{0\,n+1})$ in main-sequence stars is also known  to be  sensitive to the center physical conditions. The asymptotic theory predicts a $\langle\delta\nu_{01}\rangle=1/3\langle\delta\nu_{02}\rangle$ relationship. As it is evident in  Fig.~\ref{label_d02}.c  the ``p-mode'' spectrum for red-giant models does not follow those predictions. In particular, a large number of models have negative or very small values of $\langle\delta\nu_{01}\rangle$  independently of $\langle\delta\nu_{02}\rangle$.  Similar values of $\langle\delta\nu_{01}\rangle$ have also been found in the KEPLER data \citep{Beddingetal10} and in the oscillation spectrum of the CoRoT red-giant HR~7349 \citep{Carrier10}.  We note that the largest concentration of negative/small $\langle\delta\nu_{01}\rangle$ values   correspond to models ascending or descending  the RGB.  $\langle\delta\nu_{01}\rangle/\langle\Delta\nu\rangle$ vs. luminosity confirms this result. 

Searching for a common characteristic in the structure of these models, we find that while ascending and descending the RGB the turning points of $\ell=1$ modes are well inside the convective envelope.
 The steady He-burning models have a shallower convective envelope and the turning points of $\ell=1$ modes are inside the radiative region. Fig.~\ref{label_d01L}.a shows the variation of  $\langle\delta\nu_{01}\rangle/\langle\Delta\nu\rangle$  with the distance (in  acoustic radius $\tau(r')=\int_0^{r'} dr/c$) between the bottom of the convective zone (BCZ) and the turning point for a $\ell=1$ mode with frequency close to $\nu_{\rm max}$ ($tp_1$). In Fig.~\ref{label_d01L}.b we highlight the behaviour of $\langle\delta\nu_{01}\rangle$ for low-mass RGB models.  The scatter of $\langle\delta\nu_{01}\rangle/\langle\Delta\nu\rangle$ rapidly decreases as $\tau_{tpl_1}-\tau_{\rm BCZ}$ increases (deep convective envelope) and  $\langle\delta\nu_{01}\rangle/\langle\Delta\nu\rangle$ takes negative values for models in which $tp_1$ is well inside the convective envelope.

\section{Concluding remarks}

We presented the properties of the  theoretical spectrum of solar-like oscillations during the  RGB and core He-burning phases of red giant evolution, and analyzed the behaviour of large and small frequency separations  computed from modes well trapped in the acoustic cavity of these stars. The main results of this global overview are the following:

\begin{itemize}
\item Independently of the evolutionary state, $\ell=2$ and $\ell=3$ modes trapped in the acoustic cavity have an inertia of the same order as that of the corresponding radial mode and behave as ``p-modes'' with frequencies regularly spaced by $\langle\Delta\nu\rangle$. As a consequence, the scatter of $\ell=2$ modes in the folded \'echelle diagrams is rather small.

\item The trapping of $\ell=1$ modes in the acoustic cavity depends on the evolutionary state. While a regular pattern of dipole modes is expected in more centrally  condensed models ascending the RGB , the scatter significantly increases for core He-burning ones. Therefore the regularity  of the $\ell=1$ spectrum  could be used to discriminate between different evolutionary phases.

\item $\langle\delta\nu_{02}\rangle$  and $\langle\delta\nu_{03}\rangle$  depend almost linearly on the large separation, hence on the mean density of the model, with a slope that slightly depends on the stellar mass. 

\item $\langle\delta\nu_{01}\rangle$ seems to  reflect the distance between the $\ell=1$ turning point and the bottom of convective envelope. 
It takes negative (or small) values if $tp_1$ is well inside the convective envelope, as it occurs in models ascending or descending the RGB.

\item  The properties of our adiabatic spectra are in good agreement with those found in KEPLER data for low luminosity RGB stars  \citep{Beddingetal10};  and also with those for CoRoT-exofield red giants. In this study we predict a large scatter of dipole modes for ``Red-Clump" stars  \citep[$\Delta\nu\sim 4\mu$Hz,][]{Hekker09, Mosser10, MiglioPop09}, as well as a regular pattern of $\ell=0$ and 2 modes for all the red giants. Both features correspond quite well with the properties of the spectra found by \cite{Mosser10b} in the  CoRoT-exofield red-giants (see Fig.~\ref{label_mosser} and their Fig.~3).

\end {itemize}

\acknowledgements
J.M. and A.M. acknowledge financial support from the Prodex-ESA Contract Prodex 8 COROT (C90199) and FNRS.  The authors thank M.A. Dupret for useful suggestions concerning oscillation frequency computations.

\end{document}